\documentclass[sigconf,screen]{acmart}

\AtBeginDocument{%
  \providecommand\BibTeX{{%
    \normalfont B\kern-0.5em{\scshape i\kern-0.25em b}\kern-0.8em\TeX}}}

\copyrightyear{2021}
\acmYear{2021}
\setcopyright{acmlicensed}\acmConference[KDD '21]{Proceedings of the 27th ACM SIGKDD Conference on Knowledge Discovery and Data Mining}{August 14--18, 2021}{Virtual Event, Singapore}
\acmBooktitle{Proceedings of the 27th ACM SIGKDD Conference on Knowledge Discovery and Data Mining (KDD '21), August 14--18, 2021, Virtual Event, Singapore}
\acmPrice{15.00}
\acmDOI{10.1145/3447548.3467110}
\acmISBN{978-1-4503-8332-5/21/08}

\usepackage{makecell}

\begin{document}

\title{A Semi-Personalized System for User Cold Start Recommendation on Music Streaming Apps}

\author{Léa Briand}
%\authornote{Contact author: \href{research@deezer.com}{research@deezer.com}}
\affiliation{
  \institution{Deezer Research}
  \country{}
}
\email{research@deezer.com}

\author{Guillaume Salha-Galvan}
\affiliation{
  \institution{Deezer Research}
  \institution{LIX, \'{E}cole Polytechnique}
  \country{}
}

\author{Walid Bendada}
\affiliation{
  \institution{Deezer Research}
  \institution{LAMSADE, Univ. Paris Dauphine}
  \country{}
}

\author{Mathieu Morlon}
\affiliation{
  \institution{Deezer Research}
  \country{}
}

\author{Viet-Anh Tran}
\affiliation{
  \institution{Deezer Research}
  \country{}
}

%\author{Léa Briand$^{1}$, Guillaume Salha-Galvan$^{1,2}$, Walid Bendada$^{1,3}$, Mathieu Morlon$^{1}$, Viet-Anh Tran$^{1}$}
%\affiliation{%
%  \institution{$^{1}$ Deezer Research, Paris, France}
%  \institution{$^{2}$ LIX, \'{E}cole Polytechnique, Palaiseau, France}
%  \institution{$^{3}$ LAMSADE, Université Paris-Dauphine, Paris, France}
%  \country{}
%}
%\email{research@deezer.com}

\renewcommand{\shortauthors}{L. Briand et al.}

\newcommand{\up}[1]{\textsuperscript{#1}}

\begin{abstract}
Music streaming services heavily rely on recommender systems to improve their users' experience, by helping them navigate through a large musical catalog and discover new songs, albums or artists. However, recommending relevant and personalized content to new users, with few to no interactions with the catalog, is challenging. This is commonly referred to as the \textit{user cold start} problem. In this applied paper, we present the system recently deployed on the music streaming service Deezer to address this problem. The solution leverages a semi-personalized recommendation strategy, based on a deep neural network architecture and on a clustering of users from heterogeneous sources of information. We extensively show the practical impact of this system and its effectiveness at predicting the future musical preferences of cold start users on Deezer, through both offline and online large-scale experiments. Besides, we publicly release our code as well as anonymized usage data from our experiments. We hope that this release of industrial resources will benefit future research on user cold start~recommendation.
\end{abstract}




\maketitle
\section{Introduction}
\label{s1}
Recommender systems are essential tools for online services providing access to large catalogs, such as e-commerce websites \cite{smith2017two,wang2018billion}, and video \cite{gomez2015netflix,covington2016deep} or music \cite{jacobson2016music,schedl2018current,bendada2020carousel,epure2020multilingual} streaming services. They help users navigate through massive amounts of content, discover new products, movies or songs they might like, and are known to improve the user experience and engagement~\cite{bobadilla2013recommender,schedl2018current,mu2018survey,zhang2019deep}.

A prevalent approach to effectively recommend personalized content on these online services is \textit{collaborative filtering} which, broadly, consists in predicting the preferences of a user within a set of items by leveraging the known preferences of some similar users \cite{su2009survey,covington2016deep,schedl2018current,koren2015advances}. In particular, several recent works emphasized the empirical effectiveness of \textit{latent models} for collaborative filtering at addressing industrial-level challenges \cite{gomez2015netflix,covington2016deep,jacobson2016music,smith2017two}. In a nutshell, these models aim at directly learning vector space representations, a.k.a. \textit{embeddings}, of users and items where proximity should reflect user preferences, typically via the factorization of a user-item interaction matrix \cite{koren2009matrix,koren2015advances,he2016fast} or with neural network architectures processing usage data \cite{mongia2020deep,wang2015collaborative,covington2016deep}.

However, the performances of these models tend to significantly degrade for new users who only had few interactions with the catalog \cite{schedl2018current,lika2014facing,cao2020improving}. They might even become unsuitable for users with no interaction at all, who are absent from user-item matrices in standard algorithms \cite{lee2019melu,bobadilla2012collaborative,gope2017survey}. This is commonly referred to as the \textit{user cold start} problem \cite{schedl2018current,lee2019melu,lika2014facing,bobadilla2012collaborative,volkovs2017dropoutnet}. Yet, recommending relevant content to these new users is crucial for online services. Indeed, a new user facing low-quality recommendations might have a bad first impression and decide to stop using the service. 

In this applied paper, we present a system recently deployed on the global music streaming app Deezer\footnote{\href{https://www.deezer.com}{https://www.deezer.com}}, connecting 14 million active users from 180 countries to 73 million music tracks, to practically address this problem. The solution starts from an existing large-scale latent model for collaborative filtering, periodically trained on Deezer's \textit{warm} users (see Section \ref{s3}). It automatically integrates \textit{cold} users into the existing embedding space, by collecting heterogeneous sources of demographic and interaction information on these users at registration day, processed by a deep neural network, and by leveraging a segmentation of warm users to strengthen the final representations and provide semi-personalized recommendations to cold users \textit{by the end of their registration day}.

The proposed system is suitable for an online production use on a large-scale app such as Deezer. Throughout this paper, we extensively show its practical impact and its empirical effectiveness at predicting the future musical preferences of cold users, through both offline experiments on data extracted from Deezer and an online A/B test on the actual Deezer app. We also emphasize how this system enables us to provide more interpretable music recommendations. Last, along with this paper we publicly release our source code as well as anonymized usage data of Deezer users from our offline experiments. Besides making our results reproducible, we believe that this open-source release of industrial resources will benefit future research on user cold start, by providing to the scientific community a relevant real-world benchmark dataset to evaluate future recommender systems.

This paper is organized as follows. In Section \ref{s2}, we introduce the user cold start problem more precisely and mention previous research efforts on this topic. In Section \ref{s3}, we present the semi-personalized recommender system deployed on the Deezer app. We report and discuss our experimental setting, our data, and our results in Section \ref{s4}, and we conclude in Section \ref{s5}.

\section{Preliminaries}
\label{s2}

In this section, we provide a precise formulation of the problem we aim at addressing. We also give an overview of the existing related work. Some of the mentioned approaches will constitute relevant baselines to evaluate the effectiveness of our system.

\subsection{Problem Formulation}
\label{s2.1}

Throughout this paper, we consider a catalog of $m$ music tracks available on the music streaming app Deezer. We assume that the catalog remains fixed over time, which we later discuss. At time~$t$, Deezer gathers $n_t$ \textit{warm} users who, according to some criteria internally fixed by our data scientists, had a sufficiently large number of interactions with the catalog, e.g. enough listening sessions, to be used in the training of our recommender systems. We consider a \textit{latent model for collaborative filtering} \cite{koren2015advances,koren2009matrix,covington2016deep,mongia2020deep}. From the observed  warm user-track interactions and following the processes described in Section \ref{s3}, this model learns a vector space representation of both users and tracks. In this \textit{embedding} space, each user $i$ and track $j$ are represented by $d$-dimensional vectors (with $d \ll m$ and $d \ll n_t$), say $u_i \in \mathbb{R}^d$ and $v_j \in \mathbb{R}^d$, capturing musical preferences. To recommend relevant new tracks to the user~$i$, we leverage user-item similarity measures $f(u_i,v_j)$ in this space, encoding user-item affinities. These measures are typically based on an inner-product or a cosine similarity \cite{koren2009matrix}.  The model is updated at regular time intervals to take into account the evolution of preferences.

Everyday, new users, referred to as \textit{cold}, will register to the service. They will only have few to no interactions with the catalog during their registration day. As explained in the introduction, a straightforward inclusion of these cold users in the aforementioned latent model is unsuitable \cite{lika2014facing,bobadilla2012collaborative,gope2017survey,lee2019melu}. Waiting for them to become warm users (according to internal criteria) is also undesirable: indeed, recommending relevant content as soon as possible is crucial, as new users facing low-quality recommendations might make up their mind on this first impression and quickly stop using the service. As a consequence, we aim at addressing the following problem: \textit{given an existing latent model for collaborative filtering learning an embedding space from a set of warm users, how can we effectively include new cold users into this same space, by the end of their registration day on Deezer?}
In the remainder of this paper, we will evaluate the estimated embedding vectors of cold users by assessing their ability at predicting the future musical preferences of these users on Deezer after their registration day, through the evaluation tasks and metrics presented in Section \ref{s4}.

\subsection{Related Work}
\label{s2.2}

The user cold start problem has initiated significant research efforts over the past decade \cite{schedl2018current,lika2014facing,cao2020improving,lee2019melu,bobadilla2012collaborative,gope2017survey,volkovs2017dropoutnet,mu2018survey,zhang2019deep,smith2017two,shi2017local,kula2015metadata,lam2008addressing,lin2013addressing,bharadhwaj2019meta,zhang2019star}. In the following, we provide an overview of the most relevant work w.r.t. our approach. We refer the interested reader to some recent surveys \cite{mu2018survey,zhang2019deep,gope2017survey} for a more exhaustive review of the existing literature, and to \cite{van2013deep,mu2018survey,smith2017two,wang2018billion} for a presentation of the related \textit{item cold start} problem, which is out of our scope.

A prevalent strategy to address the user cold start problem in the total absence of usage data consists in relying on metadata related to new users, and notably on demographic information (such as the age or country of the user) collected during registration \cite{lika2014facing,kula2015metadata,lam2008addressing,mu2018survey,yanxiang2013user,fernandez2016alleviating}. In particular, various approaches aim at clustering warm users, and subsequently assign cold users to existing clusters by leveraging these metadata \cite{felicio2017multi,yanxiang2013user,mu2018survey,lika2014facing,cao2020improving,su2009survey}. Building upon these works, the model we present in Section~\ref{s3} will also leverage demographic information and incorporate a clustering component.

Besides, one can enrich such systems by explicitly asking new users to rate items from the catalog through various interview processes, leading to hybrid models based on preferences and side information \cite{gope2017survey,mu2018survey,shi2017local}. On the industry side, Netflix \cite{gomez2015netflix} and Spotify~\cite{jacobson2016music} are famous examples of services implementing such \textit{onboarding} session for new users; as explained in Section \ref{s3}, Deezer also adopted this strategy. As the inclusion of an onboarding is not always possible in production, authors of \cite{felicio2017multi} propose to use bandit algorithms to assign cold users to warm user segments, while \cite{lin2013addressing,shapira2013facebook} resort to social media data to connect similar users.

\begin{figure*}[ht]
  \centering
  \includegraphics[width=0.83\linewidth]{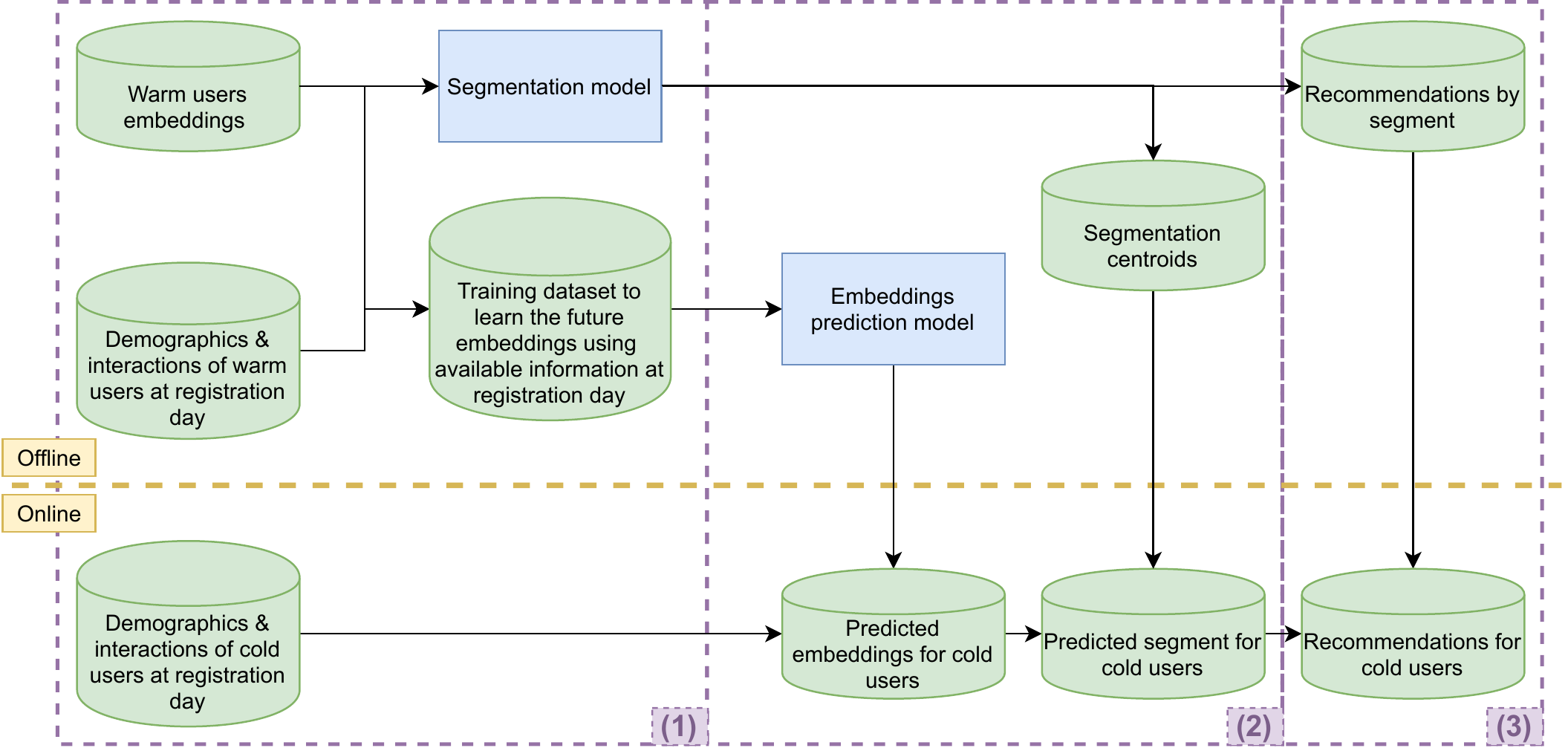}
  \caption{The semi-personalized user cold start recommendation framework available for online requests in our production environment, and described throughout Section \ref{s3}. (1)~Demographics and user-item interactions are concatenated, as described in \ref{s3.2.2}, to (2) predict a user embedding vector as in \ref{s3.2}, from warm user embeddings described in \ref{s3.1.2} and \ref{s3.1.3}. From the estimated user preferences, the new user is assigned to a segment of warm users. (3) Combining the online predicted segment with the pre-computed top items by segment, cold users benefit from semi-personalized recommendations.}
  \Description{}
\label{figOurProductionEnv}
\end{figure*}

Sometimes, cold users do have a few interactions with the catalog on their registration day. In this case, exploiting such usage signal, in addition to the aforementioned side information, can significantly improve recommendations \cite{schedl2018current,bobadilla2012collaborative,su2009survey,he2017neural}. In particular, several recent works emphasized the effectiveness of \textit{deep learning} models at dealing with such heterogeneous settings \cite{lee2019melu,bobadilla2012collaborative,volkovs2017dropoutnet,mu2018survey,zhang2019deep,covington2016deep,bharadhwaj2019meta}. 
Notably, authors of \cite{covington2016deep} explain how a scalable deep neural network, processing various user-item interactions (including watched videos, search queries...) and demographic information to learn embedding vectors, improved the YouTube recommender system. To represent interactions, they average various latent representations of watched/searched items from a same user session, allowing features to have the same dimension for each user. In the method we present in Section \ref{s3}, we will draw inspiration from their approach to preprocess the user features serving as input to our embedding model. However, a direct comparison to \cite{covington2016deep} is impossible, as no complete description nor implementation of the YouTube recommender system was made publicly available.

Among the reproducible deep learning approaches, DropoutNet \cite{volkovs2017dropoutnet} emerged as one of the most powerful latent collaborative filtering models, addressing cold start while preserving performances for warm users. This neural network takes into account usage and content data, and explicitly simulates the cold start situation during training by applying \textit{dropout} \cite{srivastava2014dropout}, alternatively to user and item embedding layers. DropoutNet relies on the assumption that data is \textit{missing at random}, with the risk of introducing biased predictions \cite{little2019statistical}. Also, it equally considers different types of positive feedback; in Section~\ref{s3}, we will furthermore consider negative feedback.
Besides, both warm and cold users' embeddings are learned during training, whereas our system will directly incorporate cold users in an existing and fixed embedding space of warm users.

Recently, \textit{meta-learning} based methods \cite{vanschoren2018meta} have also been proposed to solve the user cold start problem, with promising performances \cite{lee2019melu,zhang2019deep,bharadhwaj2019meta}. Notably, optimization-based algorithms consider each user as a learning task. From a set of global parameters ensuring an initialization of the recommender system, other local parameters are progressively updated while the user interacts with items to capture his/her preferences. In particular, authors of \cite{lee2019melu} introduce MeLU (for \textit{Meta-Learned User preference estimator}), a neural network architecture following such a meta-learning paradigm and learning preferences from the concatenation of user and item information. When new users interact with some items, then \textit{local} parameters of the neural network are updated to refine predictions for these users. In the absence of usage data, users will still be associated to an embedding vector and receive a recommended list of items, thanks to \textit{global} updates of all layers of MeLU.

Last, we point out that some research efforts also transposed recent advances in \textit{graph representation learning} \cite{hamilton2017representation} to recommender systems \cite{berg2017graph,zhang2019star,ying2018graph,wang2021graph}. Especially, \cite{zhang2019star} proposed a \textit{Stacked and Reconstructed
Graph Convolutional Network (STAR-GCN)} architecture, extending ideas from \cite{berg2017graph} to tackle the user cold start problem from bipartite user-item graphs. Although such a work deserves a mention, we will however not include it in our baselines, as the model returned memory errors in our experiments (with hundreds of thousands to millions of users) and thus did not scale to Deezer data. Nonetheless, we believe that future works on scalable graph models for user cold start, e.g. by leveraging ideas from \cite{chen2018fastgcn,salha2019degeneracy,salha2021fast}, could definitely lead to promising new methods, especially considering the recent successes of scalable graph-based recommender systems on large-scale services such as Alibaba \cite{wang2018billion} or Pinterest \cite{ying2018graph}.

\section{A Semi-Personalized System to Address User Cold Start on Deezer}
\label{s3}

In this section, we present the system deployed in 2020 on the global music streaming app Deezer to address the user cold start problem, as formulated in Section \ref{s2.1}. The architecture of our framework is summarized in Figure \ref{figOurProductionEnv}, and discussed thereafter. The proposed solution is suitable for production use on an online service.

 % to add : train model => onnx format model
 % des que nouvelle interaction, possiblement nouveau segment pédit et donc nouvelles reco

\begin{figure*}[ht]
  \centering
  \includegraphics[width=0.75\linewidth]{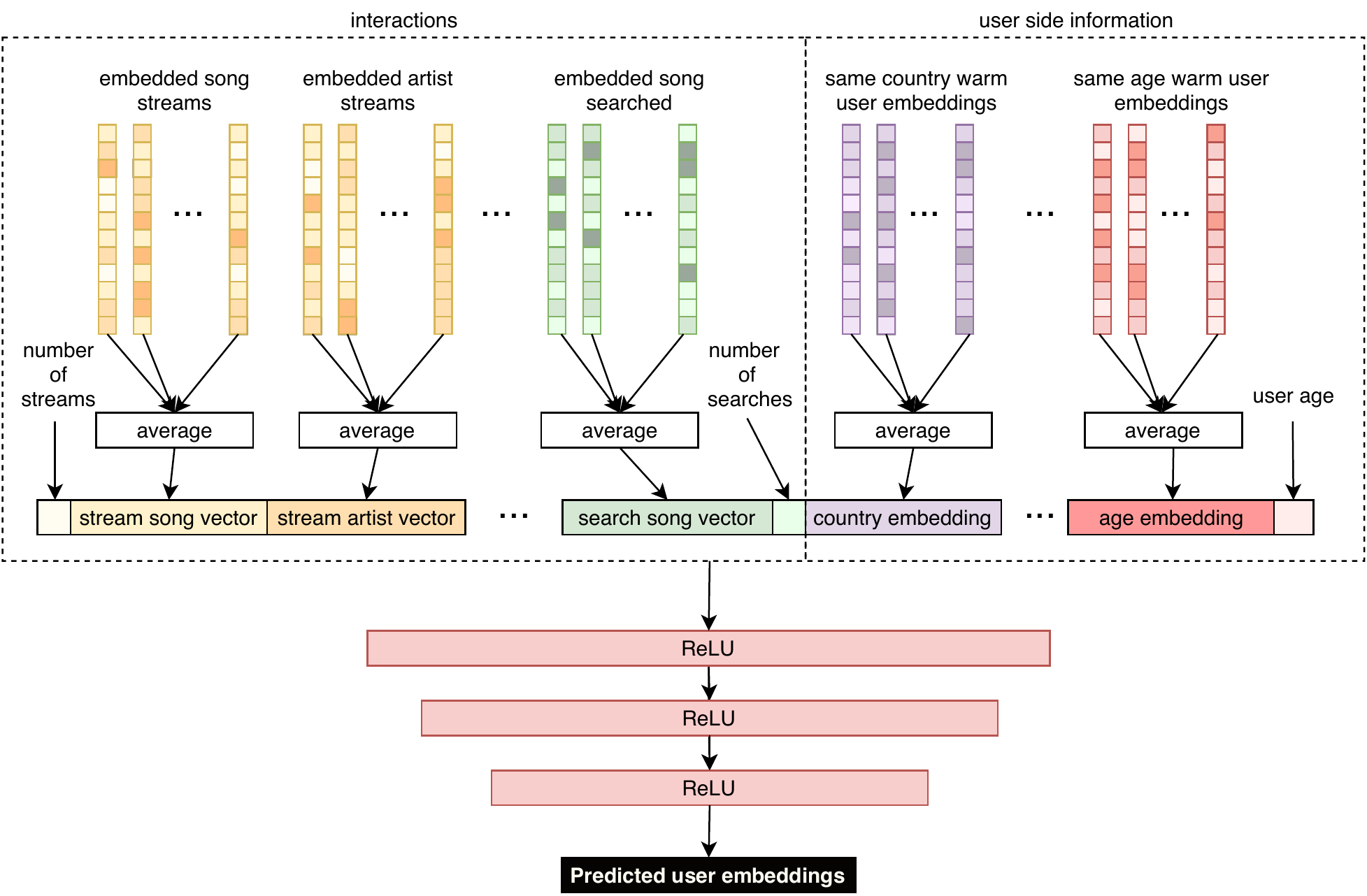}
  \caption{Prediction of user embeddings from heterogeneous data. Top left: embedding vectors of activated items in usage data, including onboarding, are aggregated as described in \ref{s3.2.2}. Top right: we enrich representations with demographic information. Bottom: after pre-processing the dense input features vector, a deep neural network model, trained as in \ref{s3.2.3}, predicts user embedding vectors in either the UT-ALS embedding space (from \ref{s3.1.1}) or the TT-SVD embedding space (from \ref{s3.1.2}).}
  \Description{Model architecture}
  \label{embeddingsPredictionFigure}
\end{figure*}

\subsection{Two Latent Models to Represent Musical Preferences of Warm Users}
\label{s3.1}

We recall that our objective is to effectively incorporate, \textit{by the end of their registration day}, a set of cold users in an existing embedding space trained on warm users. In the following, we introduce two different strategies to learn such a space on Deezer data. Although some technical details on computations are omitted for confidentiality reasons, some embedding vectors from both models will be released with this paper (see Section \ref{s4}). Moreover, experimental results on both spaces will be reported in the next section.

%\subsubsection{User tastes embeddings}

%There are diverse ways to represent users' music preferences. One way is to model them with embeddings. An embedding is a transformation of a high dimensional data into a more useful and compressed form. Ideally, an embedding is able to capture some of the semantics of the input by placing semantically similar inputs close together in the embedding space. We describe two ways of producing such user embeddings. 

\subsubsection{UT-ALS Embeddings}
\label{s3.1.1}
Latent models for collaborative filtering can approximate a preference matrix between users and items from the product of two low-rank matrices, respectively stacking up latent vector representations, a.k.a. embedding vectors, of users and items \cite{koren2015advances,su2009survey}. At Deezer, we consider a \textit{user-track (UT) interaction matrix} summarizing interactions between millions of active users and music tracks from the catalog. The \textit{affinity score} between user $i$ and music track $j$, i.e. the entry $(i,j)$ of the matrix, is computed from various signals, including the number of streams and the potential addition of the music track (or the corresponding album or artist) to a playlist of favorites.
% including the number of streams, of searches, of skips, and the potential addition of the music track (or the corresponding album or artist) to a playlist of favorite or banned tracks.
The final entry is refined in accordance with internal heuristic rules. Then, we rely on a \textit{weighted matrix factorization}, specifically by using the \textit{alternating least squares} (ALS) method \cite{koren2009matrix}, to map both users and music tracks to a joint latent space of dimension $d = 256$. These vector representations will be referred to as \textit{UT-ALS embeddings} in the remainder of this paper.

\subsubsection{TT-SVD Embeddings}
\label{s3.1.2}
 
Models inspired from word2vec \cite{mikolov2018advances} rely on the \textit{distributional hypothesis} \cite{pennington2014glove} to map items co-occurring in similar contexts to geometrically close embedding vectors. Authors of \cite{levy2014neural} show that word2vec with negative sampling implicitly factorizes a shifted \textit{pointwise mutual information} (PMI) matrix using \textit{singular value decomposition} (SVD) \cite{koren2009matrix}. In this paper, we also consider a PMI matrix, based on the co-occurrences of music tracks in diverse music collections on Deezer, such as music playlists. Then, we factorize this \textit{track-track (TT) matrix} using a distributed implementation of SVD\footnote{\href{https://github.com/criteo/Spark-RSVD}{https://github.com/criteo/Spark-RSVD}}, leading to embedding vectors of dimension $d = 128$ for each music track\footnote{Embedding dimensions of UT-ALS and TT-SVD have been optimized independently for recommendation, and are therefore different (256 vs 128). The choice of SVD vs ALS factorization is also driven by internal optimizations on Deezer data. In experiments we will simply exploit these vectors, independently, for user cold start.}. Finally, we derive embedding vectors for warm users, by averaging music track vectors over their listening history on Deezer. These vector representations, different from \textit{UT-ALS embeddings}, will be referred to as \textit{TT-SVD embeddings} in the remainder of this paper.

\subsubsection{Warm User Segmentation}
\label{s3.1.3}

On top of UT-ALS or TT-SVD user embeddings, our system also computes a segmentation of warm users, by running a $k$-means algorithm, with $k =$ 1 000 clusters/segments, in the embedding space. Each user segment is represented by its \textit{centroid} i.e. by the average of its user embedding vectors. In production, a list of the most popular music items to recommend among each warm user segment is also pre-computed.

\subsection{Predicting the Preferences of Cold Users}
\label{s3.2}

In the following we present our model, illustrated in Figure \ref{embeddingsPredictionFigure}, to integrate cold users into these embedding spaces, and subsequently predict their future musical preferences on Deezer.

\subsubsection{Overall Strategy}
\label{s3.2.1}
Firstly, we gather data from various sources, presented in \ref{s3.2.2} and referred to as \textit{input features}. They can be collected for warm users and (at least partially for) cold users. Then, we train a neural network (\ref{s3.2.3}) to map input features of warm users to their (either UT-ALS or TT-SVD) embedding vectors. Last, through a forward pass on this trained neural network, we predict embedding vectors for cold users from their input features. Cold users are therefore integrated into the existing latent space alongside warm users and each track of the catalog. This will permit computing cold user-track similarities, and even similarities between cold and warm users, which we leverage in \ref{s3.2.4} for clustering.

\subsubsection{Input features}
\label{s3.2.2}

\begin{figure}[t]
  \centering
  \includegraphics[width=0.53\linewidth]{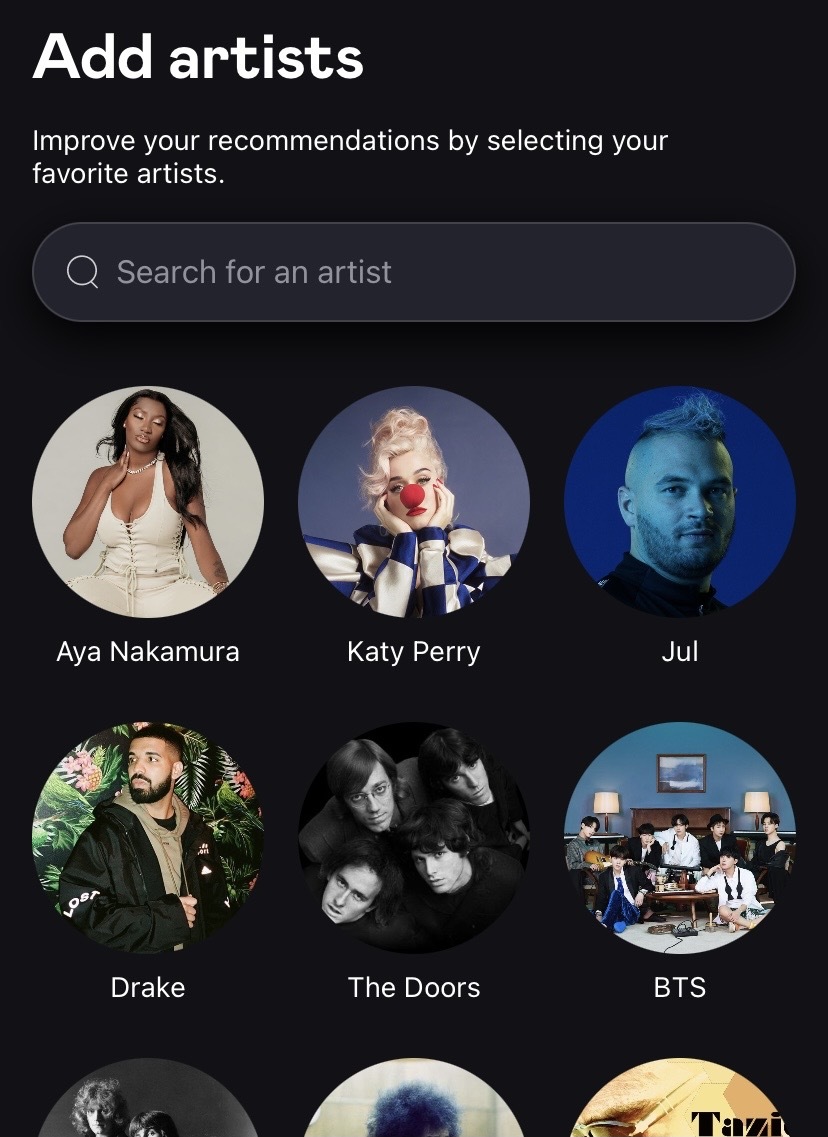}
  \caption{Overview of the \textit{onboarding} process on Deezer.}
  \Description{}
\label{onboarding}
\end{figure}

%The construction of the training dataset is crucial in our model. In our production setup, we periodically train our cold start model over a sample of 1 million warm users recently registered that have embeddings computed at the time of the training dataset computation. 
%The task is to predict the future values of wam embeddings for new users leveraging all available information at registration day. 
During registration, all users specify their age and country of origin, which we include in input features. This information is enriched with \textit{country embedding} and \textit{age embedding} vectors which, as illustrated in Figure \ref{embeddingsPredictionFigure}, are the average of embedding vectors of warm users from respectively the same country and age class. 
As hybrid models mentioned in Section \ref{s2.2}, we complement this side information with data retrieved from user-item interactions, limiting to interactions at registration day (if any). These interactions include  \textit{(positive or negative) explicit and implicit signals}, including streaming activity, searches, skips and likes. As granularity of items is crucial in music \cite{schedl2018current}, we also compute such signals at the album, artist and playlist levels, deriving embedding vectors for such music entities by averaging the relevant track embeddings (e.g. by averaging the tracks of an album, or part of the discography of an artist). Then, for each type of interaction and music entity, embedding vectors are averaged, as illustrated in Figure~\ref{embeddingsPredictionFigure}. We obtain \textit{fixed-size representations} for both demographics and interactions, i.e. independent of the number of modalities or interactions, which is crucial for scalability in a production environment. For some new users,  some (or possibly all) types of interactions at registration day may be missing; the corresponding representations are replaced arbitrarily by null vectors. To avoid such a situation, the Deezer app includes an \textit{onboarding} process for newly registered users, proposing them to add artists from various music genres to their list of favorites, as illustrated in Figure \ref{onboarding}.

\subsubsection{Model training}
\label{s3.2.3}

% Test

Fixed-size representations of demographics and interactions are concatenated to form a unique dense input vector of dimension 5139 (when considering UT-ALS embeddings) or 2579 (TT-SVD embeddings). It constitutes the input layer of a feedforward neural network, with three hidden layers of dimensions 400, 300 and 200 respectively, and an output layer of dimension $d = 256$ (when considering UT-ALS embeddings) or $128$ (TT-SVD embedding). We use \textit{rectified linear unit} (ReLU) activations at each layer except the output, followed by \textit{batch normalization} \cite{goodfellow2016deep}. We train the model on warm users, by iteratively minimizing the \textit{mean squared error} between the predicted user embeddings and their actual value in UT-ALS or TT-SVD spaces, by \textit{stochastic gradient descent} \cite{goodfellow2016deep} with a learning rate of 0.0001, batch sizes of 512, and 100 (respectively 130) epochs for TT-SVD (resp. UT-ALS). 

%\subsubsection{Hyperparameters tuning}

%For our model, we used batch of size 512 and no dropout parameter was needed in our case. Increasing the number of hidden layers did not bring any additional performance, so we kept the three ones and a number of neurons for each layer of respectively 400, 300 and 200 from the first to the last. We found optimal learning rate of 0.00005 and 0.0001, a number of epochs of 130 and 100 for the SVD and MF embeddings sets respectively. To implement the semi-personalized set up, we did a K-means with a number of 100 clusters to find with 20 maximum number of iterations on the normalized user embeddings of the training set, for each of the embeddings set.

\subsubsection{Semi-Personalization}
\label{s3.2.4}

Our model integrates cold users into the existing embedding space alongside warm users and music tracks. Therefore, one could provide \textit{fully personalized} music recommendation to each of those cold users, by retrieving the most \textit{similar} tracks for each user via an exact or an approximate \textit{nearest neighbors} search and some similarity measure. However, as we will empirically show in our experiments, such a strategy can still lead to noisy results for users with very few to no usage data. As a consequence, our system instead adopts a \textit{semi-personalized} recommendation strategy. On top of our neural network predictions, we include cold users into the pre-computed \textit{warm user segmentation}, described in Section \ref{s3.1.3}.
Specifically, each cold user is assigned to the warm cluster whose centroid is the closest w.r.t. the predicted embedding vector of this cold user. We subsequently recommend the pre-computed most popular tracks among warm users from the cluster. Our framework is summarized in Figure \ref{figOurProductionEnv}. %This framework has the advantage being updatable as soon as more interactions are observed, to refine the prediction.

\subsection{Model Deployment}
\label{s3.3}

This system is suitable for online production use on large-scale apps. At Deezer, the real-time inference service to predict user embeddings is a Golang web server, deployed in a Kubernetes cluster. The web service wraps the onnxruntime library\footnote{\href{https://www.onnxruntime.ai/}{https://www.onnxruntime.ai/}}, a fast engine for running ONNX machine learning models. It permits fast predictions of cold users' embeddings via forward passes on already trained neural networks. Models are trained offline using PyTorch, on an NVIDIA GTX 1080 GPU and an Intel Xeon Gold 6134 CPU, and then exported to ONNX format and stored on Hadoop. Embeddings of music tracks (from which we also derive embeddings of artists, albums or playlists) as well as warm segment centroids are exported weekly in tables in a Cassandra cluster, exposed via a JSON REST service. Embeddings and serialized models are weekly updated to take into account changes in the catalog and preferences, and weekly exported as well. %Each export is versioned to ensure consistency between embedding vectors, models and recommended musical items at inference time.

\section{Experimental Analysis}
\label{s4}

In the following, we evaluate the performance and impact of our system on Deezer data, through both offline and online experiments.

\subsection{Offline Evaluation}
\label{s4.1}

\subsubsection{Predicting Future Preferences of Cold Users: Evaluation Task and Metrics} 
\label{s4.1.1}
Our system permits recommending musical content to cold users. Through experiments on an offline dataset of Deezer active users, described thereafter, we evaluate to which extent the proposed recommendations at registration day would have matched the actual musical preferences of a set of users on their first month on the service. Specifically, we compute the 50 most relevant music tracks for each user of the dataset, from our model and registration day's input features (described in Section \ref{s3.2.2}). %We compare them to the actual 50 most listened tracks of each user during the next 30 days on Deezer, 
We compare them to the tracks listened by each user during their next 30 days on Deezer, using three standard recommendation metrics: the \textit{precision}, the \textit{recall}, as well as the \textit{normalized discounted cumulative gain} (NDCG) as a measure of ranking quality \cite{schedl2018current}.

\subsubsection{Dataset}
\label{s4.1.2}

For offline experiments, we extracted a dataset of 100 000 fully anonymized Deezer users. Among them, 70 000 are \textit{warm} users. They are associated to demographic information (country and self-reported age), as well as their respective UT-ALS and TT-SVD embedding vectors. These vectors correspond to those actually computed by our latent collaborative filtering models on the Deezer production system on November 1\up{st}, 2020, from \textit{millions of active warm users}. Our dataset also includes the UT-ALS and TT-SVD embedding representations of the 50 000 most popular anonymized music tracks on Deezer. 

The remaining 30 000 users are \textit{cold} users, who registered on Deezer on the first week of November 2020, and subsequently listened to at least 50 music tracks \textit{on their first month on the service (excluding registration day)}. They are split into a validation set and a test set of respectively 20 000 and 10 000 users.
For each cold user, we collected demographic information as well as the list of artists potentially selected during the onboarding session, and the lists of available streams, skips, bans, searches, additions to favorites relative to music tracks, artists, albums and playlists \textit{at registration day only}. 77\% of cold users from this dataset streamed at least once at registration day, whereas 95\% of cold users fulfilled one of the aforementioned interactions. Thus, for the remaining 5\%, only demographic information is available at registration day. Last, the dataset includes the tracks listened by cold users during their next 30 days on Deezer, among the 50 000 tracks.
%the top 50 most listened tracks of each cold user during the next 30 days on Deezer, among the 50 000 tracks.

Along with this paper, we publicly release\footnote{Data and code are available on: \href{https://github.com/deezer/semi_perso_user_cold_start }{https://github.com/deezer/semi\_perso\_user\_cold\_start}} this dataset, as well as the code corresponding to our offline experiments. Besides making our results reproducible, it will publicly provide a relevant real-world benchmark dataset to evaluate and compare future cold start models on actual (albeit anonymized) usage data. We therefore hope that this open-source release of industrial resources will favor and benefit future research and applications on user cold start problems.

\subsubsection{Models} 
\label{s4.1.3}

We report results from two versions of our system (one trained on the UT-ALS embeddings of the 70 000 warm users, and one trained on their TT-SVD embeddings) on the task presented in \ref{s4.1.1}. For each embedding space, we simultaneously evaluate:
\begin{itemize}
    \item the \textit{semi-personalized} recommendations, which are actually used in production at Deezer. In this case, the 50 recommended tracks of each user will correspond to the 50 most popular tracks of his/her \textit{user segment}, as detailed in \ref{s3.2.4} ;
    \item the \textit{fully-personalized} ones, which directly leverage the predicted embedding vectors of each cold user from the neural network. In this case, we recommend, for each cold user, the 50 \textit{nearest neighbors} music tracks w.r.t. his/her vector in the embedding space, according to a cosine similarity.
\end{itemize}

Moreover, although the main objective of this applied paper is to show the very practical impact of a deployed system and not to chase the state of the art, we also report the performances of several baseline methods as a point of comparison. Foremost, we consider a \textit{popularity-based} baseline. This chart-based method recommends the most popular songs to cold users. Besides, to motivate the need for a careful modeling of user cold start, we consider the \textit{Registration Day Streams} method, which includes cold users in the embedding in a more straightforward way. It estimates a cold user's embedding vector by averaging embedding vectors of music tracks listened at registration day (and relying on popularity in the absence of any stream), then recommends 50 tracks to each cold user via a nearest neighbors search. Moreover, a third ablation baseline, denoted \textit{Input Features Clustering} in the following, will consist in
getting rid of our neural network model and directly rely on the \textit{input features} of Section \ref{s3.2.2}. This method will also cluster all users into segments via a $k$-means, but from their stacked input features, i.e. the large vector illustrated in Figure \ref{embeddingsPredictionFigure}. Recommended tracks will correspond to the 50 most popular among warm users of each cluster; as users are no longer in the same space as tracks, we do not evaluate any fully-personalized version of this baseline.

Last, we evaluate DropoutNet \cite{volkovs2017dropoutnet} and MeLU \cite{lee2019melu}, two powerful deep learning models from the recent literature, described in Section \ref{s2}. We selected these two methods as they are simultaneously \textit{1)} among the most promising approaches, to the best of our knowledge, \textit{2)} scalable to large datasets, and \textit{3)}~publicly available online\footnote{Public implementations are available on \href{https://github.com/layer6ai-labs/DropoutNet}{https://github.com/layer6ai-labs/DropoutNet} and \href{https://github.com/hoyeoplee/MeLU}{https://github.com/hoyeoplee/MeLU} respectively.}. They process the same input features as our model, with the notable exception that DropoutNet only processes positive user-track interactions (i.e. not skips nor bans, representing approximately 12\% of all interactions in our dataset). We carefully tuned each model using the validation set. For each model, we simultaneously evaluate \textit{fully-personalized} recommendations where, as for our system, we recommend to each cold user his/her 50 most similar music tracks, as well as \textit{semi-personalized} recommendations leveraging a warm user segmentation similar to ours. We consider two variants of each model from the last two paragraphs, respectively trained on UT-ALS and TT-SVD embeddings.

%Concerning the model with only socio demographic information, the recall curve is way below all the other ones. Moreover, the more we add different kinds of interactions - adding searches to streams increases for example the  recall@10 of 5 points of percentage - the more the model performance is increased. Thus, taking into account the interactions of the new users at their registration day in addition to their socio-demographic information seems crucial, and diversify the kinds of interactions seems also to improve the segment users tastes prediction. In contrast, trying more sophisticated deep learning models did not significantly improve the model performance, even after tuning its hyperparameters.

\subsubsection{Results}
\label{s4.1.4}

\begin{table*}[!ht]

\centering
\caption{Oflline prediction of future musical preferences of Deezer cold users.}
\begin{small}
\begin{tabular}{c|ccc|ccc}
\toprule
\textbf{Methods} & \multicolumn{3}{c}{\textbf{TT-SVD Embeddings}} & \multicolumn{3}{c}{\textbf{UT-ALS Embeddings}} \\
&  \scriptsize \textbf{Precision@50 (in \%)} & \scriptsize \textbf{Recall@50 (in \%)} & \scriptsize \textbf{NDCG@50 (in \%)} & \scriptsize \textbf{Precision@50 (in \%)} & \scriptsize \textbf{Recall@50 (in \%)} & \scriptsize \textbf{NDCG@50 (in \%)}\\
\midrule
\midrule 
Popularity & 8.92 $\pm$ 0.21 & 3.01 $\pm$ 0.08 & 9.72 $\pm$ 0.20 & 8.92 $\pm$ 0.21 & 3.01 $\pm$ 0.08 & 9.72 $\pm$ 0.20 \\ 
Registration Day Streams & 9.30 $\pm$ 0.22 & 3.48 $\pm$ 0.06 & 9.73 $\pm$ 0.23 & 
16.88 $\pm$ 0.46 & 5.99 $\pm$ 0.11 & 17.72 $\pm$ 0.43 \\ 
%Registration Day Streams + pop & 11.25 $\pm$ 0.28 & 4.20 $\pm$ 0.11 & 11.89 $\pm$ 0.26 & 18.79 $\pm$ 0.38 & 6.69 $\pm$ 0.15 & 19.83 $\pm$ 0.40 \\ 
%Reg. Day Streams Semi-Pers. & 18.56 $\pm$ 0.39 & 6.60 $\pm$ 0.16 & 19.82 $\pm$ 0.43 & 19.65 $\pm$ 0.29 & 7.01 $\pm$ 0.12 & 21.42 $\pm$ 0.29 \\ 
Input Features Clustering & 8.84 $\pm$ 0.22 & 2.97 $\pm$ 0.08 & 9.75 $\pm$ 0.23 & 8.85 $\pm$ 0.22 & 2.98 $\pm$ 0.08 & 9.75 $\pm$ 0.22 \\ 
DropoutNet Full-Pers. & 10.04 $\pm$ 0.27 & 3.75 $\pm$ 0.11 & 10.46 $\pm$ 0.29 & 16.30 $\pm$ 0.50 & 5.77 $\pm$ 0.50 & 17.62 $\pm$ 0.54 \\ 
DropoutNet Semi-Pers. & 20.85 $\pm$ 0.35 & 7.55 $\pm$ 0.12 & 22.61 $\pm$ 0.36 & \textbf{19.83 $\pm$ 0.32} & \textbf{6.93 $\pm$ 0.16} & \textbf{21.55 $\pm$ 0.44}  \\
MeLU Full-Pers. & 15.00 $\pm$ 0.40 & 5.12 $\pm$ 0.17 & 16.79 $\pm$ 0.45 & 13.92 $\pm$ 0.36 & 4.71 $\pm$ 0.12 & 15.49 $\pm$ 0.39 \\ 
MeLU Semi-Pers. & 19.66 $\pm$ 0.36 & 6.87 $\pm$ 0.15 & 21.63 $\pm$ 0.40 & \textbf{19.35 $\pm$ 0.43} & \textbf{6.71 $\pm$ 0.14} & \textbf{21.33 $\pm$ 0.45} \\ 
\midrule%
\textbf{Deezer Full-Pers. (ours)} & 9.58 $\pm$ 0.18 & 3.53 $\pm$ 0.03 & 9.77 $\pm$ 0.17 & 18.50 $\pm$ 0.43 & 6.63 $\pm$ 0.10 & 20.22 $\pm$ 0.41 \\ \textbf{Deezer Semi-Pers. (ours)} & \textbf{22.75 $\pm$ 0.32} & \textbf{8.26 $\pm$ 0.15} & \textbf{24.59 $\pm$ 0.30} 

& \textbf{19.00 $\pm$ 0.42} & \textbf{6.93 $\pm$ 0.10} & 20.38 $\pm$ 0.45  \\ 
\bottomrule
\end{tabular}
\label{offlineevaluation}
\end{small}
\end{table*}

Table \ref{offlineevaluation} reports performance scores for all models on the UT-ALS and TT-SVD embeddings, along with standard deviations over ten iterations. As the \textit{Popularity} baseline is independent from embedding vectors, it obtains the same scores for the two embeddings. In both settings, \textit{Popularity} is the worst method, although performances are still fairly good for such a simple strategy.

We first focus on TT-SVD embeddings. The \textit{Registration Day Streams} baseline hardly beats \textit{Popularity} for these embeddings, which emphasizes the limits of a direct use of sparse usage data in cold start settings. On the contrary, our proposed semi-personalized system provides significant improvements (e.g. a +14.86 NDCG points increase w.r.t. \textit{Registration Day Streams}), and even reaches competitive results w.r.t. DropoutNet and MeLU. Overall, semi-personalized methods outperform their fully-personalized variants (e.g. a +13.17 precision points increase for \textit{Deezer Semi-Pers.} vs \textit{Deezer Full-Pers.}). This confirms the empirical relevance of our user segmentation strategy, and that, while the studied methods \textit{could} provide fully personalized recommendations to cold users, this strategy can lead to noisier results on real-world applications such as ours. Last, our system provides better results than a direct use of the \textit{input features} vector (\textit{Input Features Clustering} baseline), which confirms the relevance of our modeling step on top of this vector. We considered replacing our 3-layer neural network by alternative architectures between \textit{Input Features Clustering} and \textit{Deezer Semi-Pers.}, notably by a 1-layer neural network i.e. a simpler linear regression model, but did not reach comparable results.

Regarding the UT-ALS embeddings, most conclusions are consistent w.r.t. TT-SVD. \textit{Deezer Semi-Pers.} reaches quite comparable or better results w.r.t. alternatives. Among the key differences, we highlight that, while \textit{Deezer, DropoutNet and MeLU Semi-Pers} are overperforming, the fully-personalized \textit{Registration Day Streams}, \textit{DropoutNet Full-Pers} and \textit{Deezer Full-Pers} obtain significantly stronger results than on TT-SVD embeddings (e.g. a 18.50\% precision score for \textit{Deezer Full-Pers.} on UT-ALS, vs 9.58\% on TT-SVD). UT-ALS embeddings are constructed from a user-item interaction matrix, an approach appearing as better suited for nearest neighbors search.

To conclude on Table \ref{offlineevaluation}, we emphasize that, while all scores might seem relatively low, they are actually encouraging considering the intrinsic complexity of the evaluation task (predicting a few listened tracks, among 50 000). Overall, the TT-SVD version of \textit{Deezer Semi-Pers} reaches the best results. We note that the choice of @50 metrics is not restrictive. We reached consistent model rankings for @25 and @100 scores; the number of items to recommend will be a selectable parameter in our public implementation.

\begin{figure}[!t]
  \centering
  \includegraphics[width=0.8\linewidth]{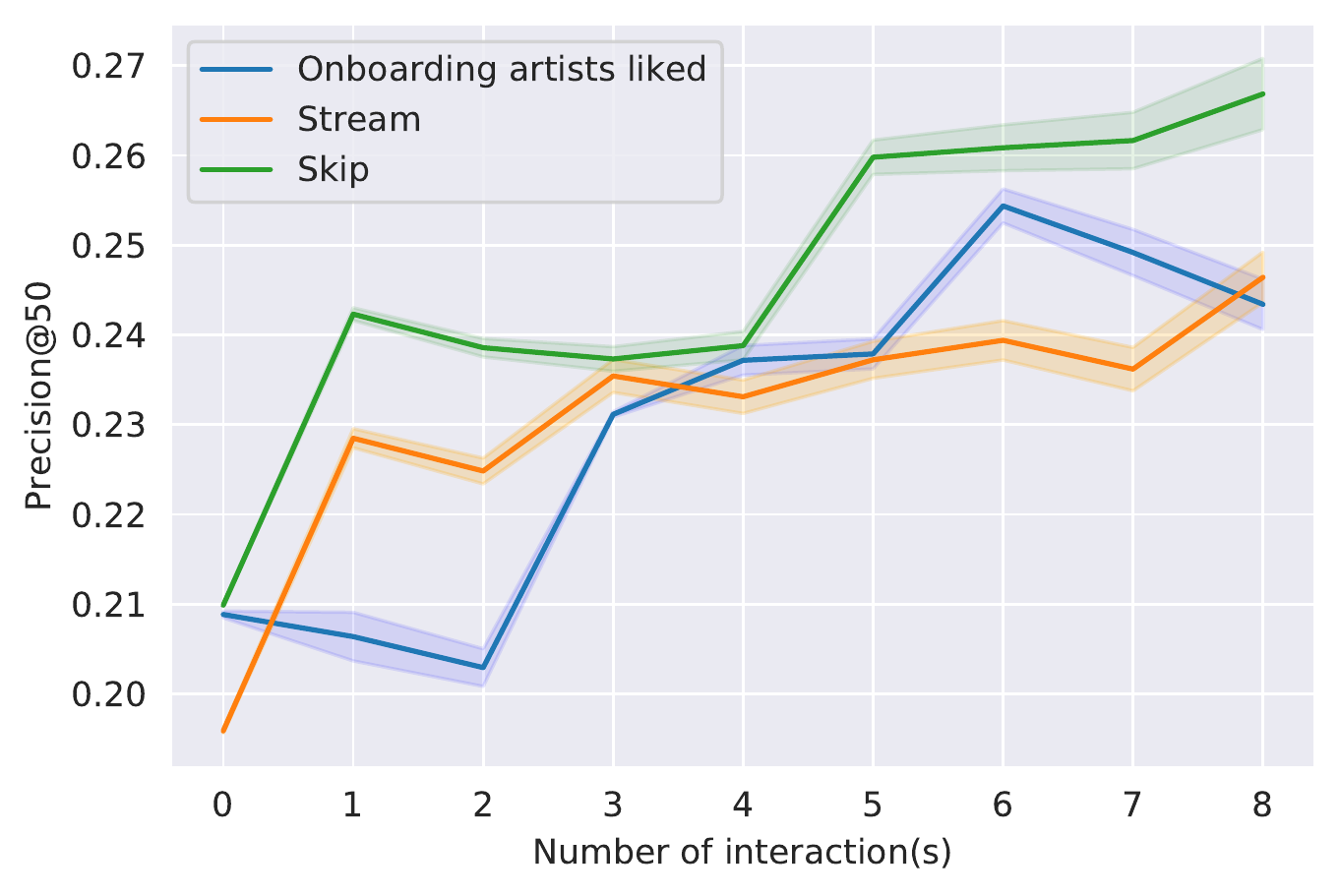}
  \caption{Precision@50 scores of Deezer Semi-Pers. depending on the number of artists liked during the onboarding session, of streams, and of skips \textit{at registration day}.}
  \Description{}
\label{interactions_analysis}
\end{figure}

To go further, Figure \ref{interactions_analysis} reports the mean precision@50 scores obtained by our semi-personalized system trained on TT-SVD embeddings, depending on available user-item interactions at registration day. 
We observe that our model returns better results for users who streamed (positive signal) or skipped (negative signal) more tracks at registration day. Liking at least three artists during the onboarding session also improves recommendations, which tends to confirm the relevance of this strategy. On the contrary, our experiments show that users without \textit{any} interaction (5\% of them, for which only demographics are available) gets a lower average precision@50 score of 17.74\%, 5.01 points below the global average.

%We want to assess the performance of our models depending on the item-consumption length of the new users. To this end, we group users into the same buckets of number of interactions, depending on which kind of interactions we study, and observe on Figure \ref{interactions_analysis} how models perform. On the left-hand side graph, users with at least one interaction see their recall@50 significantly improved compared to users without any interaction. On the right-hand side graph, we note that each kind of interactions does not equally impact the model performance. As soon as the user streams or searchs for one song, the model performance shows a great improvement, whereas users need at least 4 artists to be selected during their onboarding to show an equivalent improvement. 

%\subsubsection{Diversity Analyses}

\begin{figure}[!t]
  \centering
  \includegraphics[width=0.8\linewidth]{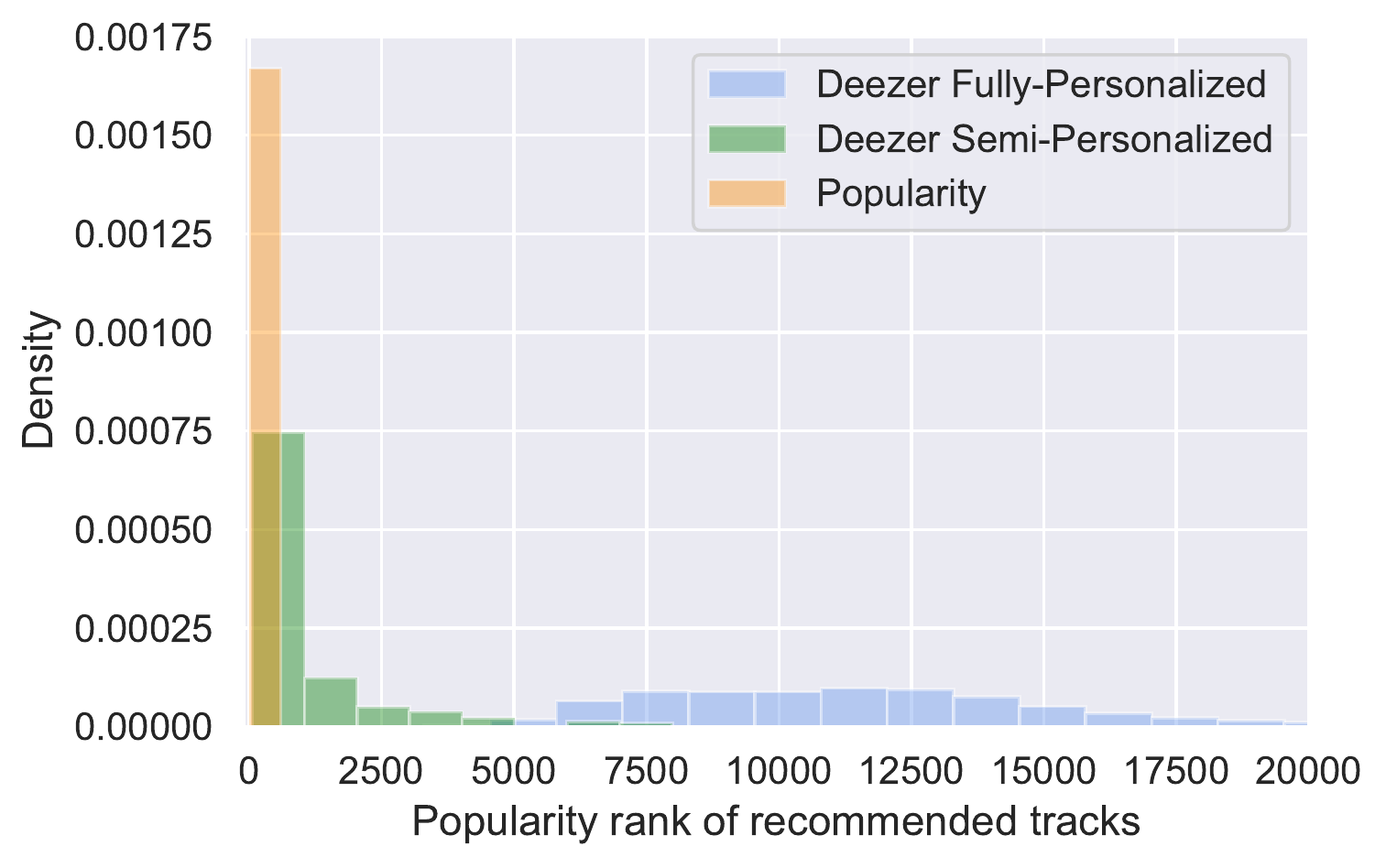}
  \caption{Distribution of music tracks recommended to cold users, by popularity rank on Deezer.}
  \label{popularity_bias_svdembed}
\end{figure}

%\subsubsection{Popularity bias}
Last, Figure \ref{popularity_bias_svdembed} reports the popularity distribution of recommended tracks from our system, trained on TT-SVD embeddings and for users from the top 5 countries on Deezer. We aim at assessing whether better performance inevitably means recommending popular tracks, which is referred to as the \textit{popularity bias} \cite{park2008long}. Our semi-personalized system stands out from the popularity baseline, and mainly recommends tracks among the 5 000 most popular in the dataset, out of 50 000. The fully-personalized variant of our system permits recommending even less \textit{mainstream} tracks, but, as we showed, this might come at the price of noisier recommendations.

%\subsubsection{Performance across music genres} The goal is to assess if the performance of our model is equally distributed among different music genres. Thanks to a mapping of the test users to their main musical genres, Figure \ref{genre_analysis} shows greater relative improvement is observed for more popular genres among our users such as for rap and rock. As SePU is strongly linked to the embeddings quality, it gives some hints on how we succeed on representing music genres across our catalog. (XXX to continue)

%\begin{figure}[h]
%  \centering
 % \includegraphics[width=0.4\textwidth]{figures/perfacrossmusicgenres.png}
%  \caption{Relative recall@50 performance of SePU + top k by segment compared to Popularity by country and age class (Pop-CA in Section \ref{models}) across different musical genres.}
 % \label{genre_analysis}
%\end{figure}

\subsection{Online Evaluation}
\label{s4.2}

\subsubsection{Online Personalization}
\label{s4.2.1}
\begin{figure}[t!]
  \centering
  \includegraphics[width=0.4\textwidth]{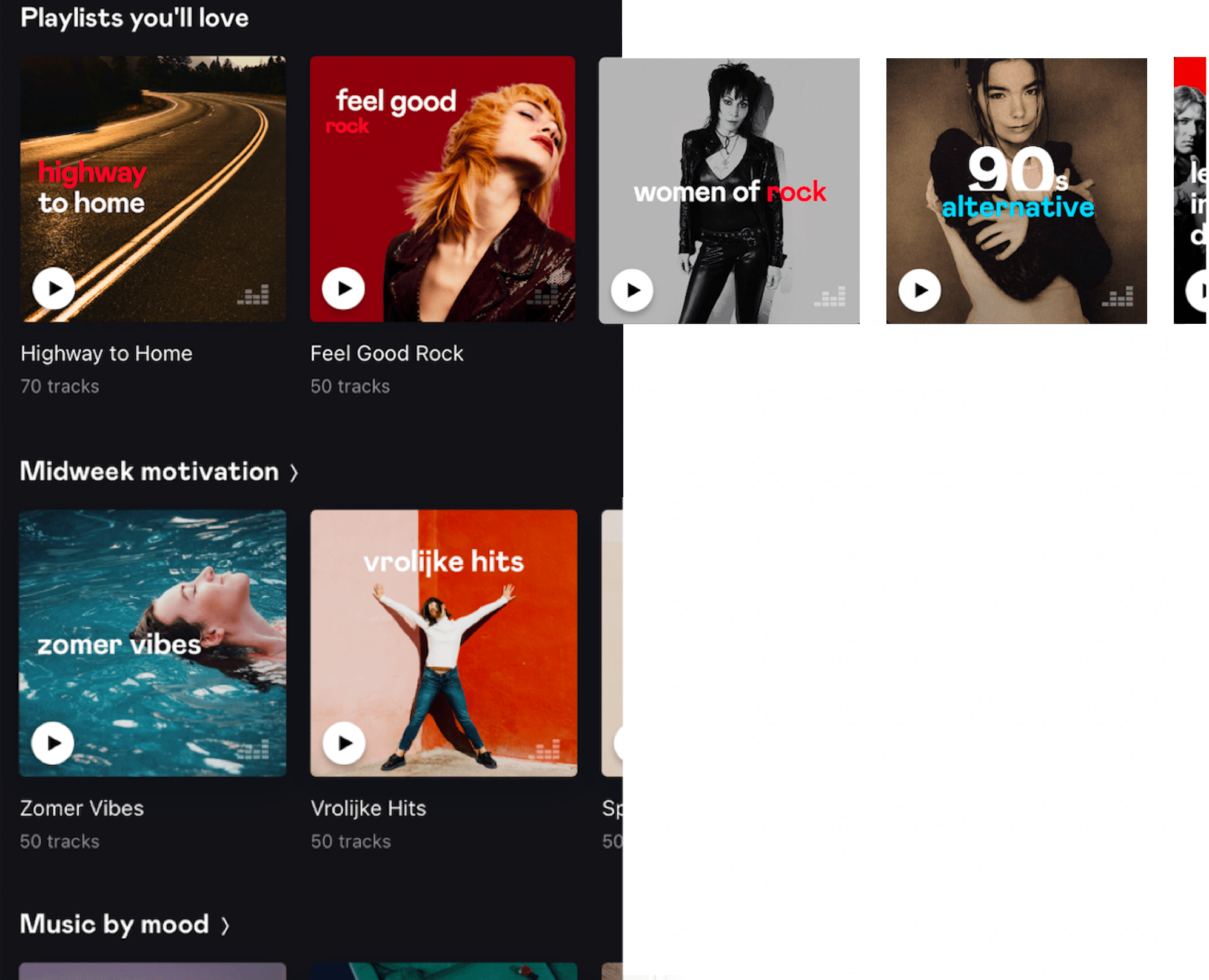}
  \caption{Personalized carousels on the Deezer app.}
  \label{carousel}
\end{figure}

In addition to these experiments on data extracted from Deezer, online tests were run to check whether our conclusions would hold on the actual Deezer app. On our homepage, we do not directly recommend music tracks, as in our offline experimental setting, but instead \textit{musical collections} such as albums and music playlists. As a consequence, our online tests will rather consist in \textit{recommending playlists to cold users}. The embedding vectors of these playlists are computed from averages of the TT-SVD embeddings of music tracks from each playlist and from internal heuristics. More precisely, Deezer displays 12 recommended playlists to each user through \textit{carousels} \cite{bendada2020carousel}, that are ranked and swipeable lists of playlists cards, as illustrated in Figure \ref{carousel}. Carousels are updated on a daily basis on the app\footnote{We provide a more detailed description to some algorithms behind dynamic carousel personalization for warm Deezer users in \cite{bendada2020carousel}.}.  All playlists were created by professional curators from Deezer, with the purpose of complying with a specific music genre, cultural area or mood.

\subsubsection{Online A/B Test Results}
\label{s4.2.2}
\begin{figure}[!t]
\includegraphics[width=0.79\linewidth]{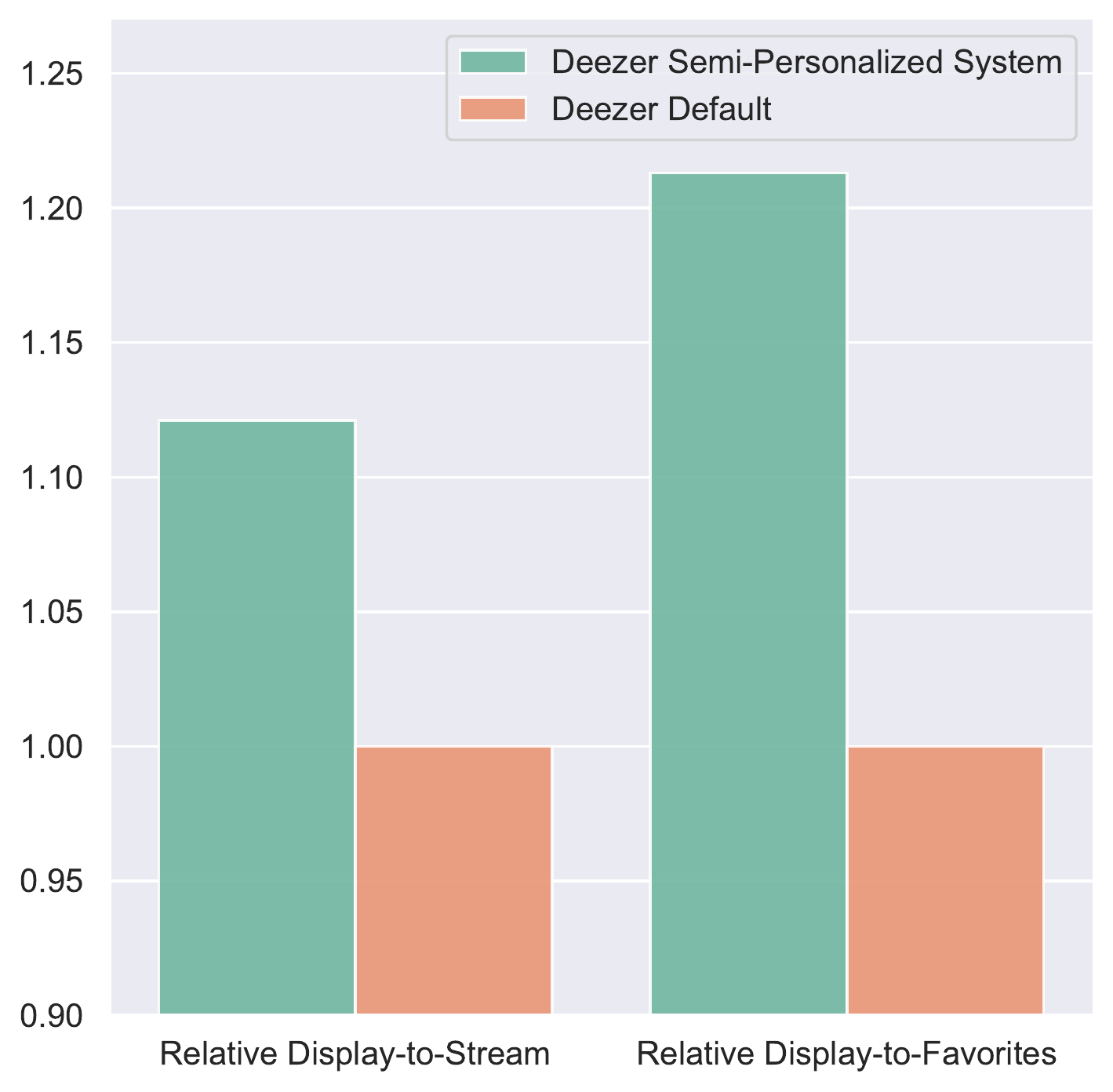}
\caption{Online A/B test: relative \textit{display-to-stream} and  \textit{display-to-favorites} rates w.r.t. internal baseline. Differences are statistically significant at the 1\% level (p-value <0.01).}
\label{abtest}
\end{figure}

A large-scale A/B test has been run for a month on Deezer in 2020, on new cold users registering during this period. Due to industrial constraints, testing all model variants in production was impossible. Also, for confidentiality reasons, we do not report the exact number of users involved in each cohort, and results are expressed in \textit{relative} terms w.r.t. the performance of \textit{Deezer Default}, a previous production system estimating cold user embeddings by countries and from internal heuristics. We observe in Figure \ref{abtest} that our new (TT-SVD-based) semi-personalized system leads to significant improvements of the relative \textit{display-to-stream} and \textit{display-to-favorite} rates, i.e. it permits selecting playlists on which cold users are more likely to click on and then to \textit{stream} the underlying content or \textit{add it to their list of favorite content}. Such results validate the relevance of our proposed system, and emphasize its practical impact on industrial-scale applications in a global music streaming app.

\subsubsection{Towards more interpretable recommendations}
\label{s4.2.3}
We underline that our semi-personalized system also enables us to provide \textit{interpretable} recommendations on Deezer. 
Indeed, carousel personalization relies on user embedding vectors that, for cold users, are linked to centroids of warm user segments. Therefore, recommended playlists can be described via the characteristics of these segments, such as the most common country or age class of warm users from each segment, or the most common music genres among their favorite artists. Table \ref{recoexamples} reports a few concrete examples of user segment descriptions, along with music playlists that were actually recommended to cold users from these segments on the Deezer app. Providing interpretable recommendations is often desirable for industrial applications, both for data scientists dealing with opaque model predictions, and for users as a way to improve their satisfaction and trust in the system \cite{afchar2020making}.

 %On Figure \ref{segmentation_analysis}, we give the example of one segment description. Such a framework helps at understanding recommendations and gives hints about what part of the pipeline can be improved. %to interpret why a user was given such recommended content. Inded, one can visualize in which kind of segment the new user is mapped. We give the examples of three different segments leading to different recommended playlists. This gives some hints why some users were recommended rap or classical editorial playlists, and helps to debug if issues are faced.

%\begin{figure}[h]
%  \centering
%  \includegraphics[width=0.6\linewidth]{figures/segmentation_analysis.png}
%  \caption{IMAGE TO UPDATE : Examples of warm user segments described by the top countries, favorite genres and playlists computed from warm users in it.}
%  \Description{}
%\label{segmentation_analysis}
%\end{figure}

\begin{table}[!t]
\centering
\caption{Examples of user segments, described by most common country, age class and music genres among favorite tracks, together with online recommended playlists.}
\begin{small}
\begin{tabular}{c|c}
\toprule
\textbf{Segment} & \textbf{Recommended Playlists on Deezer} \\
\textbf{Description} & \\ 
\midrule
\midrule
\makecell{France \\ 18-24 y.o.\\ Rap, Hip-Hop}& \raisebox{-.5\height}{\includegraphics[width=0.35\textwidth]{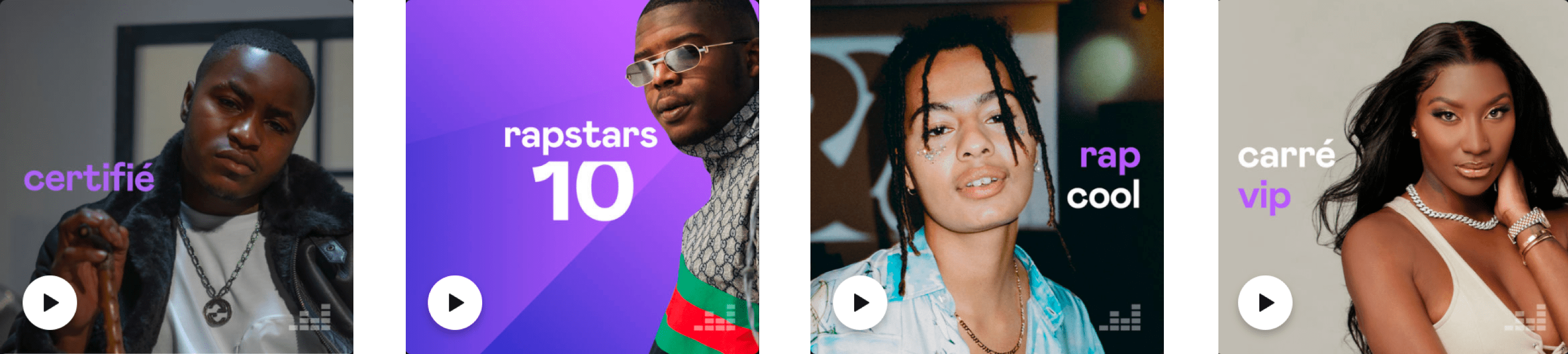}} \\
\midrule
\makecell{Brazil \\ 25-34 y.o. \\ Sertanejo}& \raisebox{-.5\height}{\includegraphics[width=0.35\textwidth]{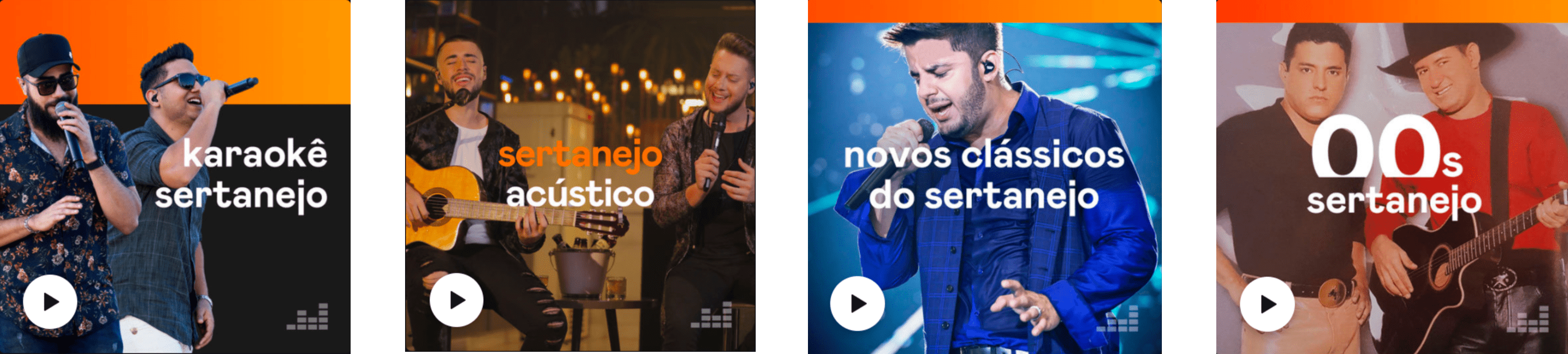}} \\
\midrule
\makecell{Germany \\ 35-49 y.o. \\ Schlager, Pop}& \raisebox{-.5\height}{\includegraphics[width=0.35\textwidth]{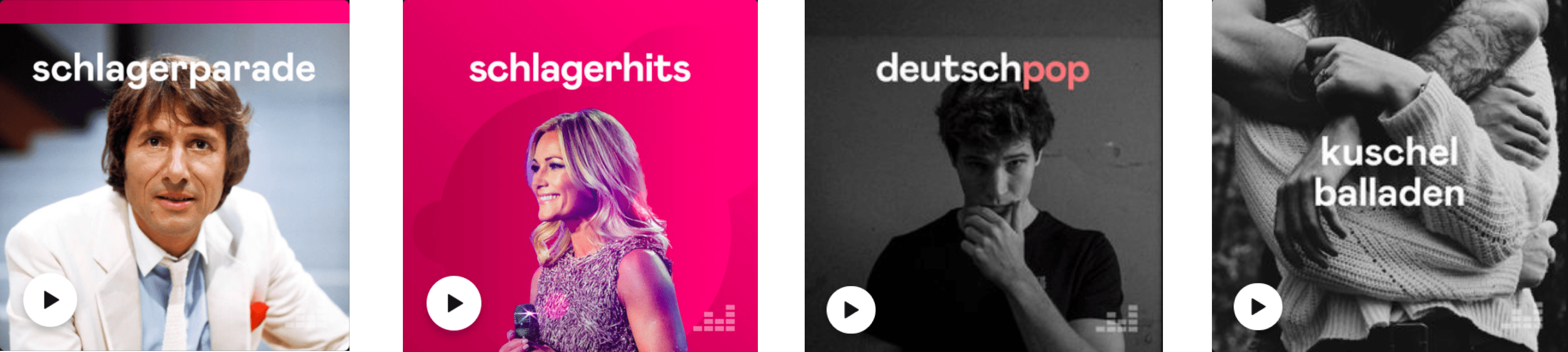}} \\
\bottomrule
\end{tabular}
\end{small}
\label{recoexamples}
\end{table}

\section{Conclusion}
\label{s5}
In this paper, we presented the semi-personalized system recently deployed in production at Deezer to address the challenging user cold start problem. We demonstrated the tangible impact of this system, through both offline and online experiments on music recommendation tasks. Moreover, although our main focus was on practical impact and not on chasing the state of the art, we also showed that our approach is competitive w.r.t. some powerful user cold start models from the recent scientific literature.
Last, along with this paper we publicly release our code, as well as the dataset used in our offline experiments, providing information on 100 000 anonymized users and their interactions with the Deezer catalog. We hope that this open-source release of industrial resources will enable future research on user cold start recommendation. For instance, in this paper we assumed that, while the number of users increases over time, the musical catalog remains fixed, which is a limit, currently addressed at Deezer via internal heuristics. Future studies on a more effective incorporation of new releases in our semi-personalized system would decidedly improve its empirical effectiveness. Besides, future studies could also consider different embedding sizes for users and items \cite{he2017neural}, or the inclusion of temporal and contextual information in our neural network. Last, as we rely on user segments, we also believe that future work on a more effective user clustering, e.g. by leveraging graph-based methods, could strengthen our system and permit providing even more refined recommendations to new users.

\bibliographystyle{ACM-Reference-Format}
\bibliography{references}

\end{document}